\documentclass{article}

\usepackage{arxiv}

\usepackage[utf8]{inputenc} 
\usepackage[T1]{fontenc}    
\usepackage{hyperref}       
\usepackage{url}            
\usepackage{booktabs}       
\usepackage{amsfonts}       
\usepackage{nicefrac}       
\usepackage{microtype}      
\usepackage{lipsum}
\usepackage{graphicx}
\usepackage{xcolor}
\usepackage{caption}
\usepackage{multirow}
\usepackage{amsmath}
\usepackage{amssymb}
\usepackage[export]{adjustbox}
\graphicspath{ {./images/} }

\DeclareMathOperator*{\argmin}{arg\,min}

\hypersetup{
pdftitle={Structure and Semantics Preserving Document Representations},
pdfauthor={Natraj Raman}
}

\title{Structure and Semantics Preserving Document Representations}

\author{
 Natraj Raman$^{1}$, Sameena Shah$^{2}$ and Manuela Veloso$^{2}$ \\
  JPMorgan AI Research\\
  $^{1}$London, UK. \\
  $^{2}$New York, USA. \\
  \texttt{first.last@jpmorgan.com} \\
}

\begin{document}
\maketitle
\begin{abstract}
Retrieving relevant documents from a corpus is typically based on the semantic similarity between the document content and query text. The inclusion of structural relationship between documents can benefit the retrieval mechanism by addressing semantic gaps. However, incorporating these relationships requires tractable mechanisms that balance structure with semantics and take advantage of the prevalent pre-train/fine-tune paradigm.  We propose here a holistic approach to learning document representations by integrating intra-document content with inter-document relations. Our deep metric learning solution analyzes the complex neighborhood structure in the relationship network to efficiently sample similar/dissimilar document pairs and defines a novel quintuplet loss function that simultaneously encourages document pairs that are semantically relevant to be closer and structurally unrelated to be far apart in the representation space. Furthermore, the separation margins between the documents are varied flexibly to encode the heterogeneity in relationship strengths. The model is fully fine-tunable and natively supports query projection during inference. We demonstrate that it outperforms competing methods on multiple datasets for document retrieval tasks.
\end{abstract}

\section{Introduction}
Document retrieval systems surface the documents that are of interest to a user based on a text query. Algorithmic solutions for document retrieval serves as a building block for several applications such as question answering~\cite{chen2017reading}, summarization~\cite{xu2020coarse}, recommendation~\cite{bhagavatula2018content}, and search and navigation~\cite{vadrevu2011scalable} and are of fundamental research interest. 

Identifying and retrieving a limited number of relevant documents is intrinsically challenging due to the long-form text, corpus size, query-document vocabulary mismatch and the asymmetry in length between the query and the document content. Modern approaches~\cite{fiorini2018best, mackenzie2018query, nogueira2019multi} follow a multi-stage cascaded ranking architecture, and their early stage retrieval typically involves representing the documents in a context-aware semantic feature space~\cite{xia2020bert}, projecting the query into the same  representation space and finally ranking the documents based on the similarity between the query representation and the document representation. 

An often overlooked aspect of this retrieval mechanism is the relationship between documents.
In real-world settings, the documents are not necessarily independent and are instead connected to each other in a network based on some shared underlying characteristics.
For instance, academic publications are related through citations, web pages by hyperlinks, clinical records with ontologies and social media via user profiles. The knowledge embedded in the corpus structure can overcome the discrepancies between a query and document due to vocabulary, granularity, implied concepts and indirect associations.
Hence computing document representations for retrieval in isolation, ignoring the valuable corpus network structure is sub-optimal~\cite{koopman2016information, cohan2020specter}. 

Existing efforts towards encoding the corpus network topology into the document representation space predominantly employ graph based architectures~\cite{chen2020graph}. A major drawback of these methods is that their transductive nature prohibits direct support for out-of-sample predictions, which are required at inference to determine the query representations. Furthermore, incorporating state-of-the-art pre-train/fine-tune contextual language model paradigm~\cite{devlin2018bert} into graph neural systems is infeasible due to resource constraints.

Metric learning~\cite{kaya2019deep} based inference is as an effective alternative to graph traversal, with support for fine-tuning and inductive inference.  
The idea here is to use a simple distance function to separate the documents based on their similarity in the representation space.
However, current attempts~\cite{reimers-2019-sentence-bert} using metric learning focus exclusively on semantic text similarity and suffer from the requirement for explicit labels to distinguish similar and dissimilar documents.
These labels are often expensive to obtain and even if available, their flat nature cannot capture the rich and complex network interactions inherent in a large corpora. Automatically determining these labels is non-trivial due to the combinatorial explosion induced by entangled neighborhood structures. In addition, existing models utilize Siamese~\cite{bertinetto2016fully} or triplet~\cite{hoffer2015deep} architectures that are not suitable for encoding the different facets of similarity.

\begin{figure*}[!tbp]
\centering
\includegraphics[width=\linewidth]{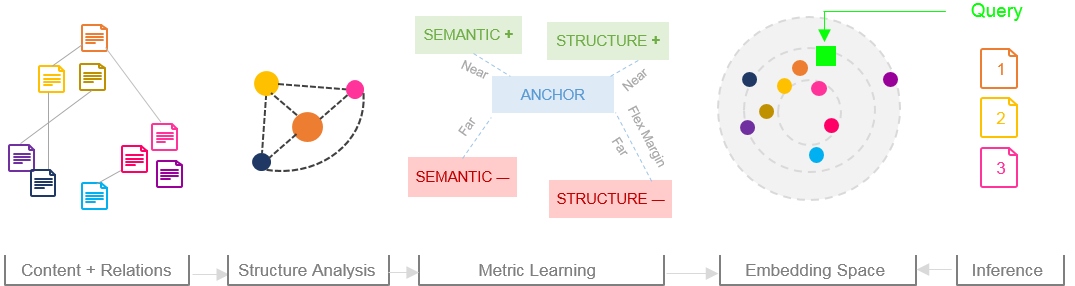}
\caption{Retrieval framework overview. Document representations are learned to preserve both the semantic content and structural relationships by efficiently mining similar/dissimilar pairs and varying the separation margins dynamically. The documents are ranked by their distance to a query projected into the same representation space.  } 
\label{fig_intro}
\end{figure*}

We address these issues here by proposing a new deep metric learning based approach for learning document representations that accounts both for intra-document content and inter-document relations.
Our solution does not require any explicit labels, and instead dynamically constructs a relative measure of similarity to separate the documents in the representation space (see Figure ~\ref{fig_intro}). 
Specifically, the corpus structure is analyzed offline to arrange the documents in an increasing order of connectedness. These ranked documents are repeatedly subdivided to sample structurally similar and dissimilar pairs while the equivalent semantic pairs are constructed from the document content. This sampling procedure covers a wide range of neighborhood in the relationship network and can scale well to large corpora. 

We extend the classical triplet loss~\cite{dong2018triplet} with a quintuplet loss function that simultaneously encourages document pairs that are semantically relevant to be closer and structurally unrelated to be far apart. This extension also addresses a key limitation in triplet loss, where the separation margins are fixed a-priori. We instead allow the margins to grow geometrically based on the the extent of structural similarity. This flexibility facilitates a relative order of separation in the representation space and enables distinguishing strong relations from weaker ones. In contrast to graph neural methods, our learned model allows the query representation to be computed seamlessly at inference. Furthermore, it supports long-form text and fine-tunes the Transformer~\cite{vaswani2017attention} language model weights adaptively during training, thereby enabling task specific customization. 

We conduct experiments on multiple publicly available datasets~\cite{cohan2020specter,sugiyama2015comprehensive,farber2018high,bird2008acl} and show that the proposed model outperforms competing methods. We also include an analysis of the learned embeddings.  Our contributions are as follows:
\begin{itemize}
  \item \textbf{Beyond Semantics}: A holistic approach to learning document representations that balances local document context with global relationship network, thereby preserving both semantics and structure.
  \item \textbf{Structure Mining}: A novel mechanism to construct similar and dissimilar pairs of documents based on a divide and conquer sampling of the neighborhood structure. 
  \item \textbf{Relative Margins}: A discriminative treatment of the representation space, encoding the nuanced relations between documents through variable units of separation.
  \item \textbf{Quintuplet Loss}: An efficient multi-input neural architecture that aggregates in parallel two different loss functions corresponding to structural and semantical facets.
  \item \textbf{Inductive and Fine-tunable}: A retrieval centric model that natively supports query projection and can be fine-tuned for task specific objectives.
\end{itemize}

In the following, Section \ref{sec_relwork} compares our work with related efforts, Section \ref{sec_model} describes the model in detail, Section \ref{sec_results} presents the results and Section \ref{sec_conclusion} summarizes our findings.

\section{Related Work}\label{sec_relwork}
Recent document retrieval approaches~\cite{mitra2018introduction, guo2020deep} are fuelled by applying deep neural networks to rank relevant documents in response to a query. Our setting is an ad-hoc top-K retrieval scenario in which there is no access to relevance judgements during training and the ranking is purely based on text similarity. Of particular interest to this paper are neural language models~\cite{devlin2018bert, sanh2019distilbert} that learn text representations by pre-training on large unsupervised corpus and allow fine-tuning on target task. While excelling at several sentence level tasks such as classification, their use of a cross-encoder makes them unsuitable for large scale semantic similarity search. A workaround is to learn embeddings that can be directly compared with a similarity metric as in ~\cite{reimers-2019-sentence-bert}. The major drawback of above models is that they focus exclusively on the content semantics, and ignore the valuable relationships between the documents. Furthermore, they are catered to sentence level inputs rather than at document level. While some models such as ~\cite{cohan2020specter} define similarity with respect to document relations, the key difference with our work is that we explicitly account both for structure and semantics in tandem. The definition of an efficient label mining procedure based on neighborhood structure further differentiates our work.

Our model modifies the conventional triplet~\cite{hoffer2015deep} network architecture with multi-instance inputs and defines a custom loss function. There have been previous efforts in using generic n-tuple inputs~\cite{chen2017beyond, zhang2019learning, chen2020adaptive} and a variety of loss functions such as contrastive loss~\cite{wang2021understanding}, triplet-center loss~\cite{he2018triplet}, lifted loss~\cite{oh2016deep}, histogram loss~\cite{ustinova2016learning}, multi-similarity loss~\cite{wang2019multi} and circle loss~\cite{sun2020circle} have been explored before. While we share with these models the general intention of designing an objective function that assigns larger weights to informative inputs, our work differs with its focus on introducing different notions of similarity rather than just improving pair selection strategy. 

Another extension in our model is the flexible variation of separation margins based on the relative strength of relationships. In \cite{ge2018deep}, a dynamic violate margin for triplet loss is formulated by constructing a class-level hierarchical tree. We differ from this by supporting a more generic graph structure. ~\cite{zhou2020ladder} proposes a graded mechanism to push inputs by distinct margins according to their relevance degree in order to construct coherent visual embeddings.  In contrast to their image mode and semantic relevance, we focus on text and the relationship structure.  

Graph representation learning~\cite{kipf2016semi, velovic2018graph, wu2019simplifying, chen2020graph} is an alternative approach to metric learning for incorporating network structure. However, its primary focus is on encoding network topology and even when text attributes are included~\cite{shen2018improved}, they are treated as side information and consequently it is not possible to fine-tune the language model for task specific customization with this approach. Notably, generalizing the embeddings to new unseen vertices is not straight-forward with these models, and in comparison our model directly supports deriving the embeddings for out-of-sample query text.

\section{Model}\label{sec_model}
This section first provides an overview of the model and then describes the components involved in detail.

\subsection{Overview}
The problem of retrieving a limited number of relevant documents corresponding to a user provided query in a self-supervised zero relevance label scenario is addressed here. A traditional retrieval mechanism that is purely based on the document content may fail to surface documents that only have a weak reference to the query and yet are highly relevant due to associational and deductive reasoning. For example, a query on \emph{coronavirus mortality} may not sufficiently match a document that discusses \emph{the benefits of vaccinations}. However, indirect associations between the concepts present in this document and the query make it highly relevant. A structured corpus could have already captured the strong relationship between documents that discuss \emph{virus mortality} and \emph{vaccinations} through topic tags or citations. We wish to utilize this auxiliary structure knowledge resource for efficient retrieval.

Our approach involves training a document representation model that encodes the corpus structure along with the content semantics into the learned document embeddings in a metric learning setting. The key idea here is to ensure that similar documents are close in the representation space while dissimilar documents are well separated. The similarity (and dissimilarity) is defined based on both a document's content as well as its relationship with other documents. The former allows meaningful comparison with a query representation while the latter bridges the above highlighted semantic gaps between a query and document during retrieval.

The representation model employs multiple instances of a deep neural network with parameter sharing to learn fixed length document embeddings. The state-of-the-art attention based Transformer~\cite{vaswani2017attention} neural architecture is adopted here. The network weights are initialized from a language model such as ~\cite{devlin2018bert} that is pre-trained on large unlabeled corpus  and the weights are adaptively tuned during training to capture target domain specific information.

The network accepts similar and dissimilar document pairs as input, one each for content semantics and corpus structure. With $K$ inputs and a training size of $N$, there are $O(N^K)$ combinations, and hence the construction of less redundant and highly informative document pairs pose a significant challenge. To tackle this, we detail an efficient pair mining procedure in the sequel that is suitable for multi-hop propagation and is based on biased sampling. Furthermore, training this deep learning network requires defining an objective function that encourages compact grouping and dispersed separation appropriately. Conventional loss functions use a margin value that is constant across the representation space to quantify the separation between similar and dissimilar document pairs. We relax this assumption and propose a novel quintuplet loss function where the margins vary dynamically based on the relationship strength for additional flexibility. 

Finally, during inference a user defined query is converted into query embeddings using the trained document representation model. These query embeddings are compared with the document embeddings to obtain similarity scores and the documents are simply ranked by these scores and returned. 

\subsection{Document Pairs Construction}
Let $\mathcal{D}=\{d_i\}_{i=1}^{N}$ be a corpus of documents. Let each document be composed of tuples of text fragments such that $d_i=(d_{i1},...,d_{is},...,d_{iS{_i}})$ where $S_i$ is the number of fragments in document $i$. The fragments may be sentences, paragraphs or any other segmentation unit including simply overlapping windows of text and are comprised of sequences of words (tokens). This formulation using document fragments that are of manageable size as inputs rather than an entire document provides direct support for long-form text. 

We assume the availability of relationship information between documents as part of the corpus structure. For example, documents that share the same topic code or documents that are referred by another document may be treated as being related. Let $A \in \mathbb{R}^{N \times N}$ be an adjacency matrix, where $A_{ij} > 0$ indicates a relationship between document $i$ and $j$. The adjacency matrix may be unweighted with $A_{ij} \in \{0,1\}, \forall{1 \leq (i,j) \leq N}$ or use real valued weights to capture the strength of the relationships.      

When training the metric learning network, for a given anchor document $i$, we wish to identify a structurally related document $\phi_i^+$ and an unrelated document $\phi_i^-$. A na\"{\i}ve random sampling from row $i$ of $A$ to identify $\phi_i^+$ and $\phi_i^-$ is inappropriate and it is important to examine the neighborhood structure present in the adjacency matrix to account for higher order proximities. Let $f:A \to \mathbb{A}$ be a function that performs link analysis~\cite{lu2011link} on $A$ and produces an intimacy matrix $\mathbb{A} \in \mathbb{R}^{N \times N}$ such that $\mathbb{A}(i,j)$ quantifies the connectivity strength between documents $i$ and $j$ after analysing the entire link structure.  We set $f$ based on the PageRank algorithm and compute $\mathbb{A}$ as
\begin{align}
\mathbb{A} = \alpha \left(\mathbb{I}-(1-\alpha)A'\right)^{-1},
\end{align}
where $\alpha \in [0,1]$ is the damping factor, $\mathbb{I}$ is the identity matrix and $A'$ is the column normalized adjacency matrix.

\begin{figure*}[!tbp]
\centering
\includegraphics[width=0.7\linewidth]{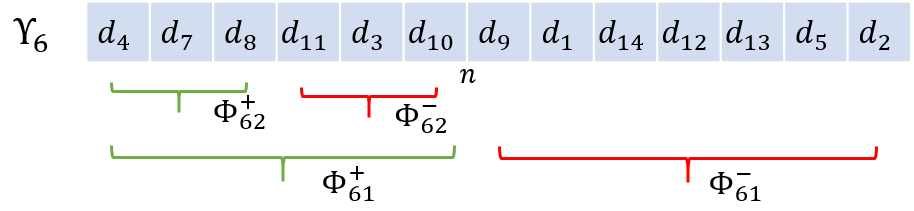}
\caption{Structural pair construction. The documents are arranged in increasing order of connectivity to the anchor document and are partitioned repeatedly to determine candidate sets of structurally related and unrelated documents. } 
\label{fig_model2}
\end{figure*}

Let $\Upsilon_i=argsort(\mathbb{A}(i,:))_{\setminus i}$ denote the sequence of documents that are in increasing order of connectivity to a document $i$, with $i$ being excluded. Let $n$ be the largest position in the sequence where $\mathbb{A}(i,\Upsilon_{i,n}) > 0$. We recursively subdivide this sequence to identify the potential set of documents that can serve as positive and negative pairs as follows:
\begin{align}
\begin{split}
    \Phi_{i1}^+ &= (\Upsilon_{i,1}....\Upsilon_{i,n}) 
    \\
    &...
    \\
    \Phi_{il}^+ &= (\Upsilon_{i,1}....\Upsilon_{i,n/2^{l-1}})
    \\
    &...
    \\
    \Phi_{iL}^+ &= (\Upsilon_{i,1}) 
\end{split}
\begin{split}
    \Phi_{i1}^- &= (\Upsilon_{i,n+1}....\Upsilon_{i,N-1}) 
    \\
    &...
    \\
    \Phi_{il}^- &= (\Upsilon_{i,(n/2^{l-1})+1}....\Upsilon_{i,n/2^{l-2}})
    \\
    &...
    \\
    \Phi_{iL}^- &= (\Upsilon_{i,2}). 
\end{split}
\end{align}

Here $\Phi_{il}$ denotes the candidate similar and dissimilar documents at level $l$ for document $i$. Figure \ref{fig_model2} provides an illustration of the partition mechanism for two levels of a target document $d_6$ with $n=6$, where $d_4$ is structurally the closest and $d_2$ is the farthest. Now $\phi_i^+$ and $\phi_i^-$ are sampled uniformly from these candidates as
\begin{align}
l &\sim \mathbb{U}[1,L] & \phi_i^+ &\sim \mathbb{U}(\Phi_{il}^+) & \phi_i^- &\sim \mathbb{U}(\Phi_{il}^-).
\end{align}

The above sampling procedure covers a wide range of the relationship network with an increased focus on the structurally related documents than the unrelated ones during partition. This inherent bias towards selecting hard triples of anchor, similar and dissimilar documents is desirable in metric learning for faster convergence. Furthermore, the explicit characterization of partition levels provides an opportunity to tailor the separation margins relative to the level from which the pairs were sampled. For example, we would expect the separation between similar and dissimilar pairs to be larger when sampled from $\Phi_{i1}$  and smaller for $\Phi_{iL}$, since the latter contains much harder examples.  

It is also necessary to identify a semantically relevant document $\psi_i^+$ and an irrelevant document $\psi_i^-$ for a given anchor $i$. We set $\psi_i^+$ to a corrupted form of the anchor document. Specifically, we replace 25\% of the tokens in the anchor with random tokens sampled from the vocabulary or a special token such as $\mathrm{[MASK]}$. Such token replacements have been hugely successful in masked language modeling ~\cite{devlin2018bert,clark2020electra} and forces the model to distinguish the tokens based on the context, thereby avoiding overfitting. The irrelevant document $\psi_i^-$ is computed through hard negative mining ~\cite{xuan2020improved} i.e. the document amongst all the other documents in a batch that is semantically closest to the anchor is the hardest negative sample and is selected as the semantically irrelevant document. 

\begin{figure*}[!tbp]
\centering
\includegraphics[width=\linewidth]{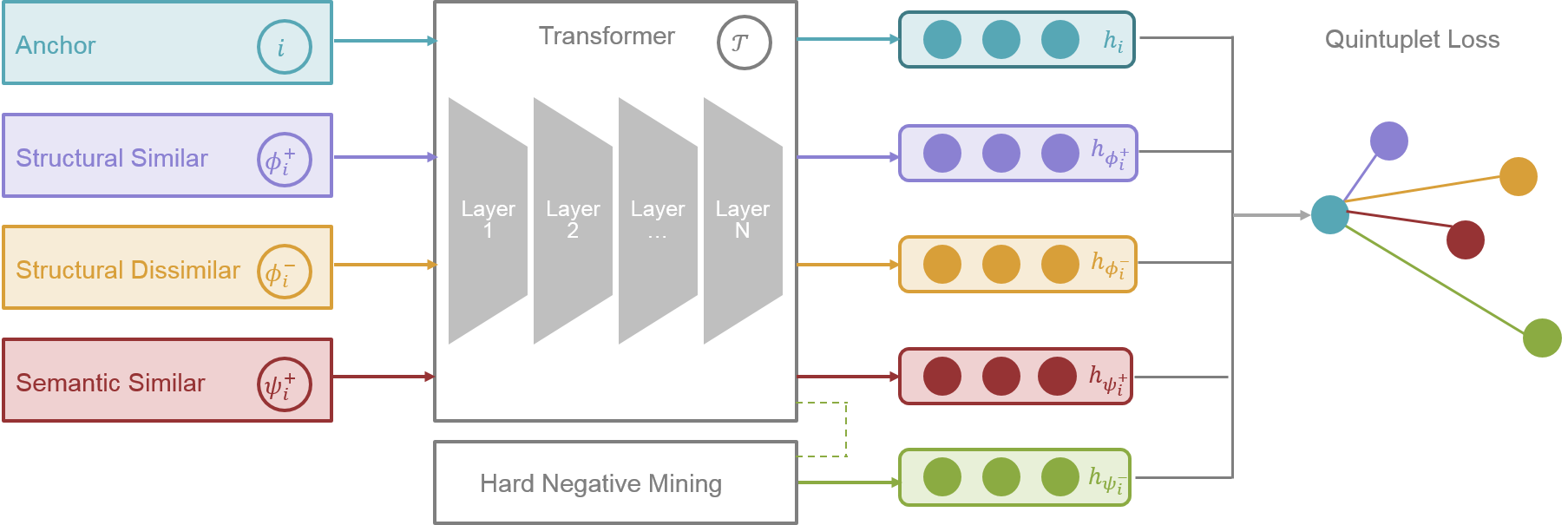}
\caption{Network Architecture. Multiple input branches are fed into a pre-trained Transformer. The loss function encourages similar documents to be closer to the anchor and dissimilar documents to be far apart.  } 
\label{fig_model1}
\end{figure*}

\subsection{Quintuplet Loss}
The document embeddings are learned by an extension to the classical Siamese~\cite{bertinetto2016fully} and triplet~\cite{hoffer2015deep} networks, with four different input branches that share the same neural architecture. The outputs from these branches are tied together using a common loss function. The four input branches correspond to an anchor $i$, structurally similar document $\phi_i^+$, structurally dissimilar document $\phi_i^-$ and semantically similar document $\psi_i^+$. Note that the semantically dissimilar document $\psi_i^-$ is determined online dynamically as outlined above, and hence does not require a separate input branch. Figure ~\ref{fig_model1} illustrates the network design.

The Transformer neural architecture uses multiple self-attention and feed-forward layers to produce the aggregate representation of an input sequence. Let $\mathcal{T}: W \to h$ be the Transformer function that accepts a sequence of tokens $W$ and produces a fixed length vector $h \in \mathbb{R}^E$ and $\delta : (a,b) \to \mathbb{R}^+$ be a distance function. Since the documents are segmented into text fragments, it is necessary to first sample the fragment index and use it as input to the transformer network. Hence the input embeddings are computed as follows:
\begin{align}
\begin{split}
    s_1 &\sim \mathbb{U}[1,S_i] \\
    s_2 &\sim \mathbb{U}[1,S_{\phi_i^+}] \\
    s_3 &\sim \mathbb{U}[1,S_{\phi_i^-}] \\
    s_4 &\sim \mathbb{U}[1,S_{\psi_i^+}] \\
    &
\end{split}
\begin{split}
    h_i &= \mathcal{T}(d_{is_1}) \\
    h_{\phi_i^+} &= \mathcal{T}(d_{\phi_i^+s_2}) \\
    h_{\phi_i^-} &= \mathcal{T}(d_{\phi_i^-s_3}) \\
    h_{\psi_i^+} &= \mathcal{T}(d_{\psi_i^+s_4}) \\
    h_{\psi_i^-} &= \argmin_{j \in \mathcal{B}} \delta(h_i, h_j) 
\end{split}
\end{align}

Thus we have five sets of embeddings $(h_i,h_{\phi_i^+},h_{\phi_i^-},h_{\psi_i^+},h_{\psi_i^-})$, with the last entry being determined online by comparing an anchor embedding with all the other embeddings in a mini batch $\mathcal{B}$ and picking the closest embedding. The loss function must encourage the distance from the anchor embeddings to a similar document embedding to be less than that to a dissimilar document embedding during training. Furthermore, we need to incorporate a margin term to specify the minimum separation distance. 

Let $m_\phi$ and $m_\psi$ be real-valued margin hyper-parameters corresponding to structure and semantic components respectively and $l_i$ be the partition level as in Equation (3). The quintuplet loss function $\mathcal{L}$ is defined as 
\begin{align}
\begin{split}
    \mathcal{L}_{\phi} &= \sum_{i=1}^N max \left\{\left(\delta(h_i,h_{\phi_i^+})-\delta(h_i,h_{\phi_i^-})+\frac{m_\phi}{l_i}\right)  ,0\right\} \\
    \mathcal{L}_{\psi} &= \sum_{i=1}^N max \left\{\left(\delta(h_i,h_{\psi_i^+})-\delta(h_i,h_{\psi_i^-})+m_\psi\right)  ,0\right\} \\
    \mathcal{L} &= (1-\gamma)\mathcal{L}_{\phi} + \gamma \mathcal{L}_{\psi},
\end{split}
\end{align}
where $\mathcal{L}_{\phi}$ is the structural relation loss, $\mathcal{L}_{\psi}$ is the semantic relevance loss and $\gamma \in [0,1]$ is a hyper-parameter that controls the relative importance between these two losses. It is important to note that the margins for structure is scaled by the inverse of partition level, thereby varying the separation distance based on the connectivity strength of the similar and dissimilar documents to the anchor. This flexibility overcomes the limitation in traditional triplet loss where all the dissimilar points are pushed away by an equal margin.

\subsection{Inference}
Given a query sequence of tokens $q$ during inference, its embedding vector is computed dynamically as $h_q=\mathcal{T}(q)$. This vector is compared with the embedding vectors corresponding to all the fragments in all the documents of the corpus for ranked retrieval. Let $\Delta : (i,j) \to [0,1]$ be a bounded similarity function and $r_{is}=\Delta(h_q,h_{is})$ be the query similarity score with fragment $s$ of document $i$. We can now determine the set of top $K$ similarity scores $R_i$ from $r_{is}, \forall s=1...S_i$.  The aggregated similarity score for document $i$ is then computed as
\begin{align}
\begin{split}
    score_i &= \sum_{k=1}^K w_k R_{ik},
\end{split}
\end{align}
where $w_k$ is the weight for position $k$. We set $w_k=e^{-\omega k}$, where $\omega$ is a hyper-parameter. Intuitively, the similarity value at position 1 will contribute more to the overall document score than say the value at position 5.  Finally, these scores are used to perform the best ranking of documents in the corpus.

\section{Experiments}\label{sec_results}
We first detail the different datasets and experiment settings used for evaluation, and then present the results and discussion.

\subsection {Datasets}
The SciDocs~\cite{cohan2020specter} dataset contains a subset of scientific papers available in Semantic Scholar. We treat the paper abstracts as the text content, while the citation graph serves as the relationship network. The dataset provides ground-truths for a recommendation task, which can be used for evaluating document retrieval. In particular, it collects data from user clickthrough logs of an academic search engine to construct a set of similar papers for a query paper title. We use this clickthrough data purely for testing our model on query retrieval and do not use it for training. 

The other datasets used for evaluation are: (a) arXivCS~\cite{farber2018high}, a document corpus from computer science publications in arXiv.org, with the citation structure mined from the \TeX files, (b) Scholarly~\cite{sugiyama2015comprehensive}, which contains publications from ACM digital library and (c) ACL-ARC~\cite{bird2008acl}, a corpus of publications about computational linguistics. Since there are no explicit abstract tags in these datasets, we use the first 20 sentences of a paper as the text content.  Similar to SciDocs, we use the references for relation structure. There is no explicit ground-truth available with these datasets for evaluating retrieval. Hence, we use a citation context sentence (only after the first 20 sentences to ensure out-of-samples) as the query and expect to find content from the cited paper in the retrieved results.

\begin{table*}[!tbp]
\caption{Comparison of retrieval results for recommendation task in  SciDocs~\cite{cohan2020specter} dataset. }
\centering
\begin{tabular}{|l|r|r|}
 \hline
 Model & R@5  & R@10  \\
  \hline
Word2Vec & 15.7 & 22.5 \\
BERT + No Fine-tune  & 16.2 & 24.6 \\
DistilBERT + No Fine-tune & 15.0 & 23.1 \\
BERT + LangModel Fine-tune & 20.6 & 29.7 \\
DistilBERT + LangModel Fine-tune & 16.9 & 27.0 \\
SciBERT  & 23.0 & 29.9 \\
SGC & 20.5 & 39.2 \\
SentBERT & 30.5  & 43.7 \\

\hline
BERT + Structure and Semantics  & 31.6 & 46.1 \\
DistilBERT + Structure and Semantics & 29.2 & 46.6 \\
SciBERT + Structure and Semantics & 35.9 & 53.9 \\
\hline
\end{tabular}
\label{tab_scidocs}
\end{table*}

\begin{table*}[!tbp]
\caption{Comparison of retrieval results for citation task in  Scholarly~\cite{sugiyama2015comprehensive}, arXivCS~\cite{farber2018high} and ACL-ARC~\cite{bird2008acl} datasets. }
\centering
\begin{tabular}{|l|r|r|r|r|r|r|}
 \hline
 \multicolumn{1}{|l|}{}&\multicolumn{2}{c|}{Scholarly}&\multicolumn{2}{c|}{arXivCS}&\multicolumn{2}{c|}{ACL-ARC} \\
 \cline{2-3} \cline{4-5} \cline{6-7}
 Model & R@5  & R@10 & R@5  & R@10 & R@5  & R@10 \\
  \hline
BERT + LangModel Fine-tune & 13.0 & 16.0 & 9.5 & 12.1 & 5.6 & 8.3 \\
DistilBERT + LangModel Fine-tune & 12.5 & 16.1 & 8.9 & 11.6 & 5.3 & 7.8 \\
SciBERT  & 12.8 & 16.8 & 11.2 & 15.0 & 6.2 & 8.7 \\
SentBERT & 26.8  & 31.7 & 21.5 & 26.2 & 14.0 & 17.9 \\
\hline
BERT + Structure and Semantics  & 28.3 & 35.1 & 22.5 & 28.4 & 14.3 & 19.7 \\
DistilBERT + Structure and Semantics & 26.4 & 33.3 & 21.8 & 27.9 & 15.4 & 20.8 \\
SciBERT + Structure and Semantics & 35.7 & 43.3 & 29.6 & 36.4 & 20.8 & 27.6 \\
\hline
\end{tabular}
\label{tab_otherds}
\end{table*}

\begin{table*}[!tbp]
\caption{Comparison of retrieval results for self prediction task across the datasets. }
\centering
\begin{tabular}{|l|r|r|r|r|}
 \hline
 Model & SciDocs  & Scholarly & arXivCS  & ACL-ARC \\
  \hline
BERT + LangModel Fine-tune & 56.5 & 10.9 & 10.5 & 10.4 \\
DistilBERT + LangModel Fine-tune & 52.2 & 10.8 & 10.9 & 11.4 \\
SciBERT  & 52.7 & 13.9 & 14.6 & 14.9 \\
SentBERT & 78.3 & 20.7 & 24.4 & 22.4 \\
\hline
BERT + Structure and Semantics  & 76.3 & 23.3 & 24.5 & 21.5 \\
DistilBERT + Structure and Semantics & 75.2 & 21.3 & 23.6 & 21.4 \\
SciBERT + Structure and Semantics & 83.9 & 27.8 & 31.9 & 31.7 \\
\hline
\end{tabular}
\label{tab_titlecomp}
\end{table*}

\subsection {Evaluation Framework}
We compare the performance of our retrieval model with several competing methods. The aggregated word vectors from Word2Vec~\cite{mikolov2013distributed} is an effective baseline model for evaluating distributed vector representations in the semantic feature space.  For context-aware representations, we include the standard BERT~\cite{devlin2018bert} and DistilBERT~\cite{sanh2019distilbert} models with average embeddings of all the tokens, which tend to perform better than using $\mathrm{CLS}$ token output for semantic similarity tasks. We also fine-tune these models with text from our target datasets using the standard language model training objective to evaluate the effects of exposing task specific content. Additionally, we compare with SciBERT~\cite{beltagy2019scibert}, a language model based on BERT but pre-trained on a large corpus of scientific text, to assess the impact of structure inclusion on different types of Transformer models. For comparisons with equivalent loss functions, we use SentBERT~\cite{reimers-2019-sentence-bert}, a state-of-the-art model that uses triples of anchor, semantically similar and dissimilar sentences to learn the embeddings. Finally, we also compare with a state-of-the-art deep graph representation algorithm SGC~\cite{wu2019simplifying}, in which graph convolutions are applied over text features based on the neighbourhood structure.  

The hyper-parameters are set as follows: structure separation margin $m_\phi$ to $2$, semantic separation margin $m_\psi$ to $0.5$, relative loss value $\gamma$ to $0.5$, score factor $\omega$ to 0.05 and damping factor $\alpha$ to 0.15. The document fragments are created using overlapping windows of text with a sequence length of $128$.  The training procedure uses Adam optimizer with a learning rate of  $5e-5$ and an epsilon of $1e-8$. The distance function $\delta$ is set to Euclidean, while cosine similarity is used during inference. We train our models on an 8 GPU NVIDIA Tesla V100 instance with a batch size of $24$ for $1$ epoch.

\subsection {Results}

\begin{figure*}[!tbp]
\centering
\includegraphics[width=\linewidth]{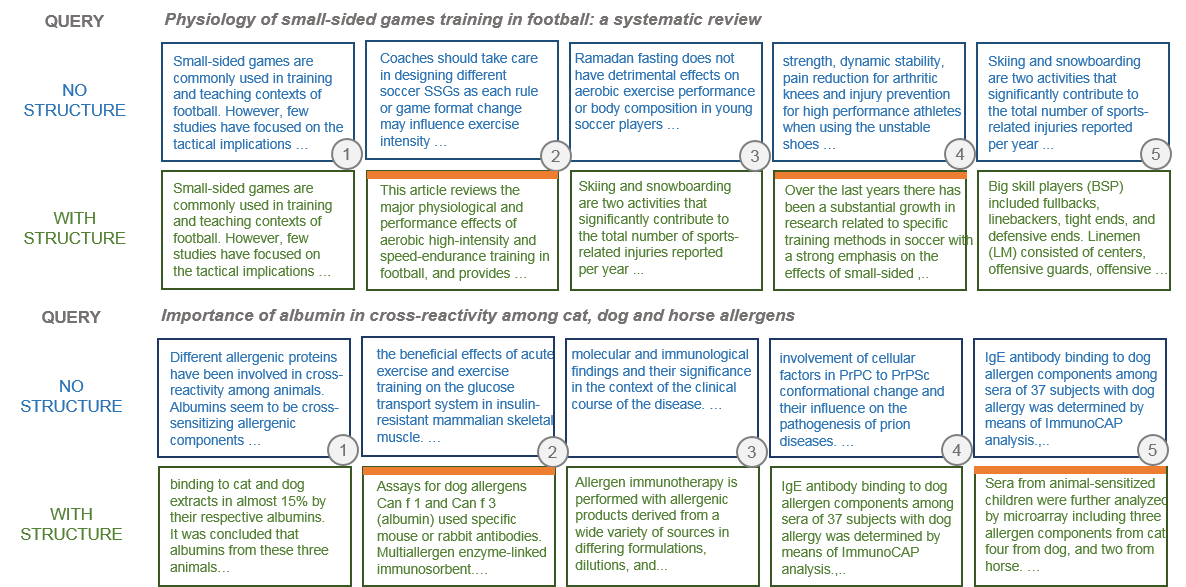}
\caption{Qualitative comparison of retrieval results. The returned documents are relevant to the query, with some overlap between the two models. However, the model that incorporates corpus structure returns the documents that are truly relevant to the user (highlighted in orange) .} 
\label{fig_exp1}
\end{figure*}

The comparison results for the SciDocs dataset is furnished in Table ~\ref{tab_scidocs}. We use Recall@K metric, which measures the percentage of relevant documents from ground-truth being returned in the top K results. For this dataset, the query is the paper title as searched by the user, while the relevant documents are those papers from the search results that were clicked by the user and marked as ground-truth. While the Word2Vec and vanilla BERT/DistilBERT models perform poorly, fine-tuning the latter with the abstracts seem to improve the recall rate. The SciBERT model  performs better than BERT/DistilBERT owing to it being already trained on scientific data. By including structure information, the SGC model improves over the above models. However, its lack of support for fine-tuning inhibits its ability to perform effective semantic similarity. The closest match to our model in performance is SentBERT, which also uses metric learning in a fine-tune paradigm albeit considering only semantic similarity. The importance of adding structure with semantics is evident from the nearly $2\%$ improvement offered by our models when compared with SentBERT, and an even larger difference for the SciBERT variant.

\begin{table*}[!tbp]
\centering
\begin{tabular}{cc}
\includegraphics[width=.45\linewidth,valign=t]{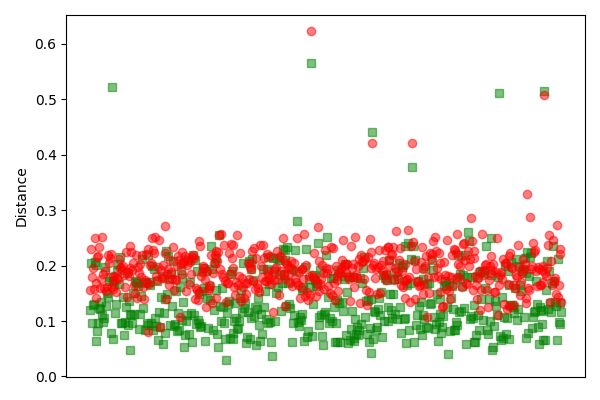} &
\includegraphics[width=.45\linewidth,valign=t]{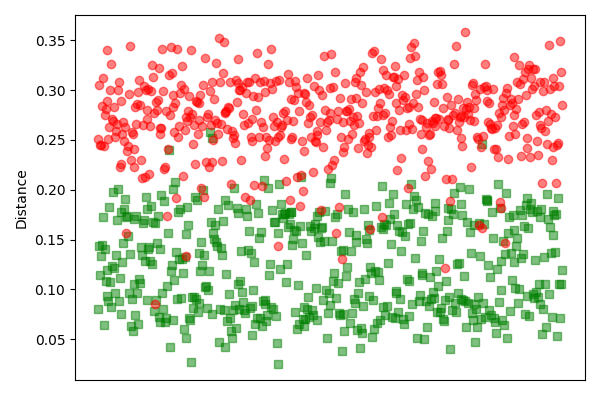} \\
(a) No Structure & (b) With Structure
\end{tabular}
\captionof{figure}{Embedding space comparison. Distance between an anchor and structurally related document embedding is plotted as green square, while an anchor and structurally unrelated document is in red circle. The model learned by including the corpus structure shows vivid separation. }
\label{fig_exp2}
\end{table*}

\begin{table*}[!tbp]
\centering
\begin{tabular}{cc}
\includegraphics[width=.45\linewidth,valign=t]{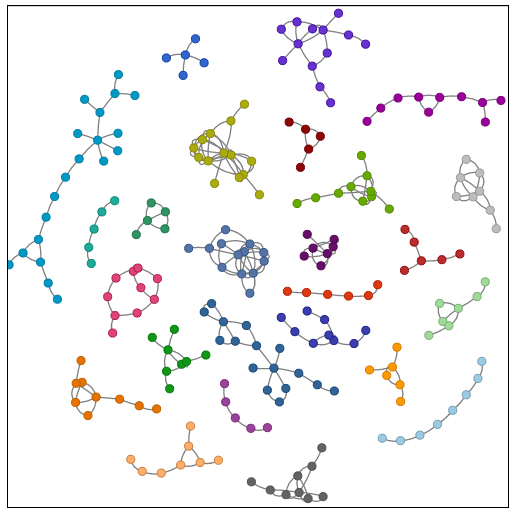} &
\includegraphics[width=.45\linewidth,valign=t]{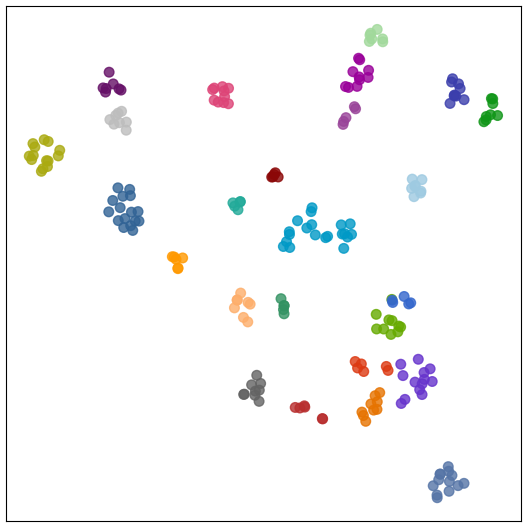} \\
(a) Ground-truth Document Relationships & (b) 2D Projected Embeddings 
\end{tabular}
\captionof{figure}{Structure preserving representation. The relationship network is preserved in the representation space with the document nodes in a subgraph appearing close together in the t-SNE projection of learned document embeddings. }
\label{fig_exp3}
\end{table*}

\begin{figure}[!tbp]
\centering
\includegraphics[width=8cm]{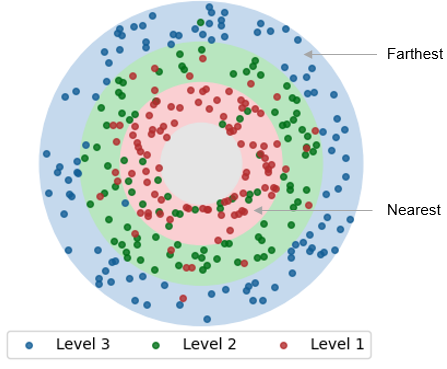}
\caption{Relative margin effect. Embedding distances increase progressively based on the relationship strength, with structurally related Level 1 pairs having smaller distance rank (Nearest) than unrelated Level 3 pairs that have large rank (Farthest).   } 
\label{fig_expvarmargin}
\end{figure}

Table ~\ref{tab_otherds} shows the quantitative comparisons for the other three datasets on the citation prediction task. The behavior seen for SciDocs is also consistently repeated for Scholarly, arXivCS and ACL-ARC datasets. Note that the citation context is used as query for these datasets, and it is possible that the abstract in the paper being cited does not sufficiently match the query text in the citation context. This is reflected in the overall low recall rates across the models. However, the relative difference between the models are still a useful assessment measure and the furnished results confirm the utility of including both semantics and structure rather than just the semantics.

We also use sample sentences from papers (which were never seen during training) as query and assess whether the model returns the corresponding paper during retrieval at the top position. For SciDocs we use the title as query, while for the other datasets we use a sentence from the paper body. Unlike the recommendation and citation tasks, this self prediction task intuitively relies only on the intra-document content and hence is a good measure of evaluating any loss of performance  in semantic matching due to the incorporation of structure. The results are presented in Table ~\ref{tab_titlecomp} and yet again our model mostly out-performs others. While our SciBERT model comfortably improves over the rest, our BERT/DistilBERT variant does not consistently exceed SentBERT. This is unsurprising since SentBERT's primary strength is on semantic similarity, which aligns well with this particular task. 

Sample queries and their corresponding top 5 retrievals for SciDocs dataset are displayed in Figure ~\ref{fig_exp1}. Snippets of the document fragments returned by a semantics only model is shown in blue while those returned by our structure + semantics model is shown in green. A qualitative assessment substantiates their relevance to the query.  The overlapping results between the two models is unsurprising considering their shared objective on semantic similarity. However, the model that incorporates structure seems more likely to return the truly relevant documents, as marked in the ground-truths.  

\subsection {Representation Space Analysis}
In order to validate whether the structure information is reflected in the learned document embeddings, we plot in Figure ~\ref{fig_exp2} the distances between documents that are structurally related in green and the documents that are unrelated in red. It can be seen in the left figure that for the model that ignores structure, the related and unrelated documents have a similar distribution for the distances. In contrast, the right figure using our model appears less cluttered with evident separation. This indicates that the learned embeddings have smaller distances between structurally similar documents and larger distances for dissimilar documents.

Additionally, we visually verify the structure preserving nature of the model in Figure ~\ref{fig_exp3}. The left side figure shows the ground-truth relationship structure for a few sample documents in the SciDocs dataset. For illustration purposes, the documents are colored by the subgraph they belong to. On the right side, the 2D embeddings of these documents obtained using t-SNE~\cite{van2008visualizing} projection is plotted. It can be seen that the documents within the same subgraph are cohesive together in the learned representation space. This cohesiveness of structurally related documents is beneficial because even if a relevant document does not sufficiently match a query, by virtue of its proximity to another connected document that contains the query terms, the relevant document's probability of being retrieved increases. Such a behavior is essential to surface indirect associations and logical deductions. 

We also investigate the effect of varying the structural margins in a relative manner rather than fixing them to a constant value. Given an anchor and a structurally similar document, we identify a set of documents with varying degrees of dissimilarity to the anchor and compare the embedding distances between these dissimilar pairs and the similar pair. The variation in dissimilarity levels is due to the differences in relationships - e.g. the anchor may have a direct connection to the similar document (Level 0), one-hop connection to a dissimilar document (Level 1), a two-hop connection to another (Level 2) and no direct connection at all to the third dissimilar document (Level 3). We should expect the distance of the anchor to the documents progressively increase from one level to another. 

Figure ~\ref{fig_expvarmargin} displays a radial plot of the differences in distances between Level 0 pair and the pairs at other levels for sample documents from SciDocs dataset. The level of a point is differentiated by color while its position depends on the distance rank. For instance, a point lying on the outer circle indicates a larger distance and on the inner circle the smallest distance. Intuitively, the Level 1 points should all lie on the inner circle and the Level 3 points on the outer circle, since a closer relationship implies nearer distance and vice-versa. This desirable behavior can be observed with a majority of Level 1 (red) points lying on the inner circle and Level 3 (blue) points placed on the outer circle. Even when not conforming to the ideal case, the errors are largely restricted to a single level misplacement. The introduction of flexible margins facilitates in imposing such an implicit order of separation between the documents and can thus effectively capture the nuances in  relations.    
 
\begin{figure}[!tbp]
\centering
\includegraphics[width=8cm]{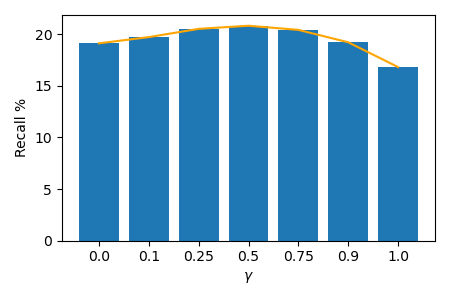}
\caption{Sensitivity of loss parameter $\gamma$ on retrieval.  } 
\label{fig_expwt}
\end{figure}

\begin{figure}[!tbp]
\centering
\includegraphics[width=8cm]{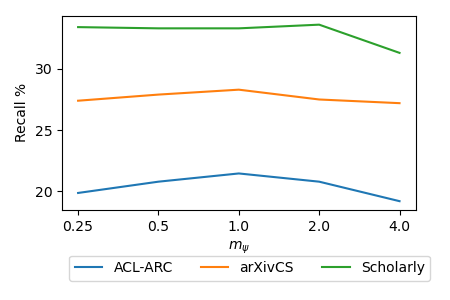}
\caption{Impact of semantic margin parameter $m_{\psi}$ on retrieval for different datasets.  } 
\label{fig_expsemmargin}
\end{figure}

\subsection {Parameter Sensitivity Analysis}
We also perform sensitivity analysis on the various parameters to gauge the importance of tuning them. In Figure ~\ref{fig_expwt} the recall rate of ACL-ARC dataset for different choices of $\gamma$ is displayed. By design, the loss is concentrated on the structure component if $\gamma=0$, while a value of $1$ focuses on the semantics. It can be observed that extreme values that degenerate to consider only one of these components is not ideal, with the optimal values lying over a wide range between $0.25$ and $0.75$. Interestingly, ignoring the structure has a relatively greater penalty. This is possibly because the semantic component is already accommodated in a partial manner due to the use of pre-trained language model weights. 

The parameter $m_{\psi}$ influences the separation margin enforced by the semantic component during training. The recall performance of our DistilBERT model for different choices of this parameter corresponding to various datasets is plotted in Figure ~\ref{fig_expsemmargin}. While excessively large values result in a drop in performance, the results are stable between values of $0.5$ and $1.5$, indicating a broad range of options.

\section{Conclusion}\label{sec_conclusion}
We introduced a new representation learning mechanism that integrates structural relations with semantic information to enrich the document embeddings.  
The sampling procedure based on link structure analysis helps traverse even complex neighbourhoods and the quintuplet loss function offers a flexible balance between the structure and semantic facets.
The variation of separation margins in accord with the relationship strength results in a more coherent representation space.
Our experiments illustrate the utility of this model for retrieving documents that are relevant to a query.
In future, we wish to leverage deep graph embeddings and extend the model to support multi-modal inputs.

\section*{Acknowledgments}{This paper was prepared for information purposes by the Artificial Intelligence Research group of JPMorgan Chase \& Co and its affiliates (“JP Morgan”), and is not a product of the Research Department of JP Morgan.  J.P. Morgan makes no representation and warranty whatsoever and disclaims all liability for the completeness, accuracy or reliability of the information contained herein. This document is not intended as investment research or investment advice, or a recommendation, offer or solicitation for the purchase or sale of any security, financial instrument, financial product or service, or to be used in any way for evaluating the merits of participating in any transaction, and shall not constitute a solicitation under any jurisdiction or to any person, if such solicitation under such jurisdiction or to such person would be unlawful. © 2021 JP Morgan Chase \& Co. All rights reserved.}

\bibliographystyle{unsrt}  
\bibliography{references}

\end{document}